# Progress towards an accurate determination of the Boltzmann constant by Doppler spectroscopy

C Lemarchand, M Triki, B Darquié, Ch J Bordé, C Chardonnet and C Daussy[1]

Laboratoire de Physique des Lasers, UMR 7538 CNRS, Université Paris 13, 99 av. J.-B. Clément, 93430 Villetaneuse, France

E-mail: christophe.daussy@univ-paris13.fr

**Abstract**

In this paper, we present significant progress performed on an experiment dedicated to the determination of the Boltzmann constant, $k_B$, by accurately measuring the Doppler absorption profile of a line in a gas of ammonia at thermal equilibrium. This optical method based on the first principles of statistical mechanics is an alternative to the acoustical method which has led to the unique determination of $k_B$ published by the CODATA with a relative accuracy of $1.7 \times 10^{-6}$. We report on the first measurement of the Boltzmann constant by laser spectroscopy with a statistical uncertainty below 10 ppm, more specifically 6.4 ppm. This progress results from improvements in the detection method and in the statistical treatment of the data. In addition, we have recorded the hyperfine structure of the probed $\nu_2$ saQ(6,3) rovibrational line of ammonia by saturation spectroscopy and thus determine very precisely the induced 4.36 (2) ppm broadening of the absorption linewidth. We also show that, in our well chosen experimental conditions, saturation effects have a negligible impact on the linewidth. Finally, we draw the route to future developments for an absolute determination of $k_B$ with an accuracy of a few ppm.

## 1. Introduction

A renewed interest in the Boltzmann constant is related to the possible redefinition of the International System of Units (SI) [1-12]. A new definition of the kelvin would fix the value of the Boltzmann constant to a value determined by The Committee on Data for Science and Technology (CODATA). Currently, the value of the Boltzmann constant $k_B$ essentially relies on a single acoustic gas thermometry experiment by Moldover et al. published in 1988 [13, 14] (to avoid confusion with $k$ generally reserved to the wave vector, we denote the Boltzmann constant by $k_B$ throughout this paper). The current relative uncertainty on $k_B$ is $1.7 \times 10^{-6}$ [15]. Besides some projects following Moldover's approach [16-19], an alternative approach based on the virial expansion of the Clausius-Mossotti equation and measurement of the permittivity of helium is very promising [20-25]. Since 2004 we have developed a new approach based on laser spectroscopy in which $k_B$ is determined by a frequency measurement. The principle [26, 27] consists in recording the Doppler profile of a well-isolated absorption line of an atomic or molecular gas in thermal equilibrium in a cell. This profile reflects the Maxwell-Boltzmann distribution of velocities along the laser beam axis. In a first experiment we have demonstrated the potential of this new approach [28-30], on an ammonia rovibrational line. We were soon followed by at least four other groups who started similar experiments on $CO_2$, $H_2O$, acetylene and rubidium [31-35].

---

[1] Author to whom any correspondence should be addressed.

In this paper, we present the large thermostat used to control the gas temperature and the new spectrometer developed to record the $\nu_2$ saQ(6,3) rovibrational line of ammonia both by linear and saturated absorption spectroscopy. We report on the first measurement of the Boltzmann constant by laser spectroscopy with a statistical uncertainty below 10 ppm and give a first evaluation of the uncertainty budget, which shows that the effect of the hyperfine structure of the probed line needs to be taken into account.

**2. The experimental setup**

The principle of the experiment consists in recording the linear absorption of a rovibrational ammonia line in the 10 μm spectral region, the ammonia gas being at thermal equilibrium in a cell. The width of such a line is dominated by the Doppler width due to the molecular velocity distribution along the probe laser beam. A complete analysis of the line shape which can take into account collisional effects (including pressure broadening and the Lamb-Dicke-Mossbauer (LDM) narrowing), hyperfine structure, saturation of the molecular transition, optical depth, etc. leads to a determination of the Doppler width and thus to $k_B$. The e-fold half-width of the Doppler profile, $\Delta\omega_D$, is given by:

$$\frac{\Delta\omega_D}{\omega_0} = \sqrt{\frac{2k_B T}{mc^2}}$$, where $\omega_0$ is the angular frequency of the molecular line, $c$ is the velocity of light, $T$ is the temperature of the gas and $m$ is the molecular mass. Uncertainty on $k_B$ is limited by that on $mc^2/h$ (directly deduced from atom interferometry experiments [36-38]), on atomic mass ratios measured in ion traps [39], on $h$ the Planck constant, on $T$, and on the ratio $\frac{\Delta\omega_D}{\omega_0}$.

The probed line is the $\nu_2$ saQ(6,3) rovibrational line of the ammonia molecule $^{14}NH_3$ at the frequency $\nu = 28953693.9\,(1)$ MHz. This molecule was chosen for two main reasons: a strong absorption band in the 8-12 μm spectral region of the ultra-stable spectrometer that we have developed for several years and a well-isolated Doppler line to avoid any overlap with neighbouring lines [40]. The experiment requires a fine control of: (i) the laser intensity sent in the absorption cell; (ii) the laser frequency which is tuned over a large frequency range to record the linear absorption spectrum; (iii) the temperature of the gas which has to be measured during the experiment.

*2.1. The spectrometer*

The spectrometer (Figure 1) is based on a $CO_2$ laser source which operates in the 8-12 μm range. For this experiment, important issues are frequency stability, frequency tunability and intensity stability of the laser system. The laser frequency stabilization scheme is described in reference [41]: a sideband generated with a tunable electro-optic modulator (EOM) is stabilized on an $OsO_4$ saturated absorption line detected on the transmission of a 1.6-m long Fabry-Perot cavity. The laser spectral width measured by the beat note between two independent lasers is smaller than 10 Hz and the laser exhibits frequency instability of 0.1 Hz ($3\times10^{-15}$) for a 100 s integration time.

Since its tunability is limited to 100 MHz, our $CO_2$ laser source is coupled to a second EOM which generates two sidebands *SB-* and *SB+* of respective frequencies $\nu_{SB+} = \nu_L + \nu_{EOM}$ and $\nu_{SB-} = \nu_L - \nu_{EOM}$ on both sides of the fixed laser frequency, $\nu_L$. The frequency $\nu_{EOM}$ is tunable from 8 to 18 GHz. The intensity ratio between these two sidebands and the laser carrier is about $10^{-4}$. After the EOM, a grid polarizer attenuates the carrier by a factor 200 but not the sidebands which are cross-polarized.

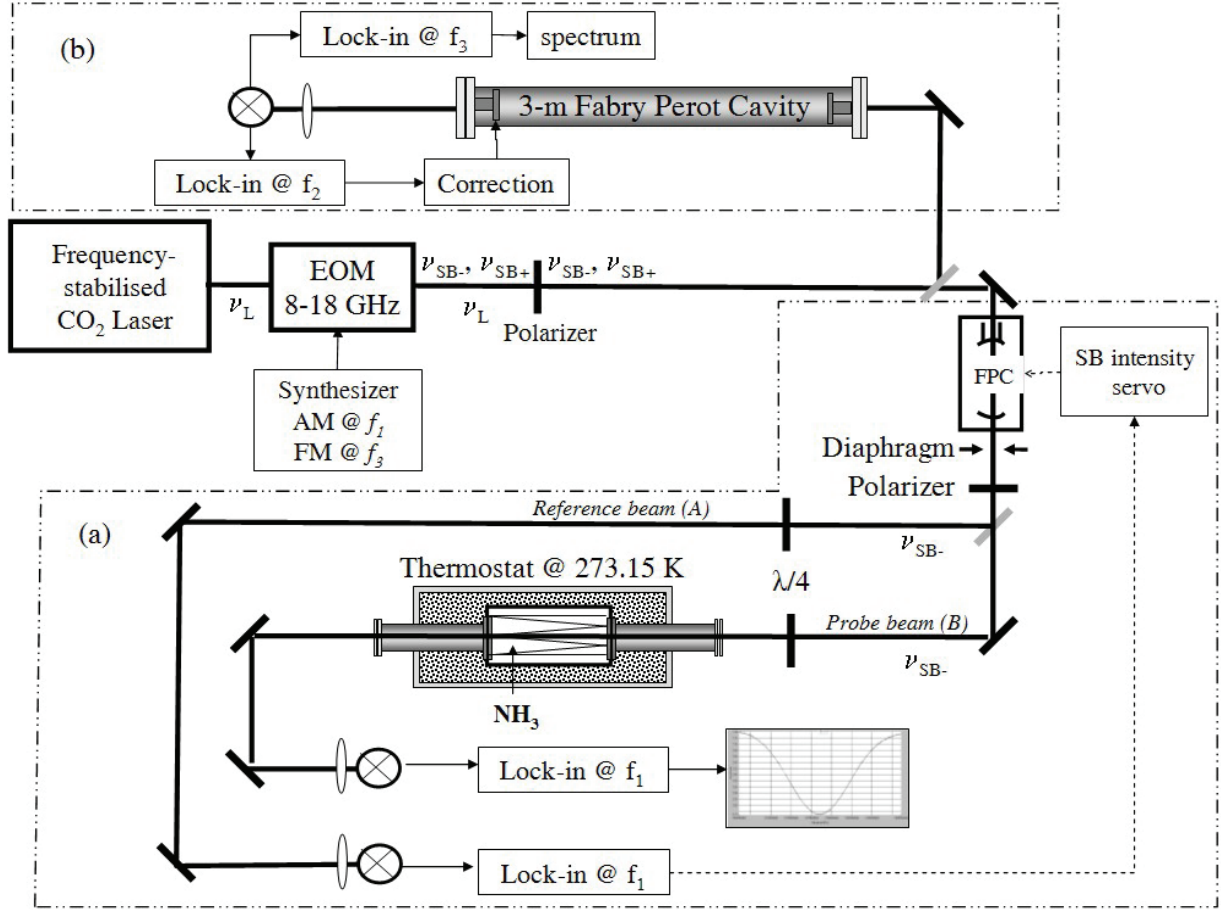

**Figure 1.** Experimental setup for (a) linear absorption spectroscopy and (b) saturated absorption spectroscopy (AM: amplitude modulation, FM: frequency modulation, EOM: electro-optic modulator, FPC: Fabry Perot cavity, *SB*: sideband, Lock-in: lock-in amplifier).

Figure 1.b represents the saturated absorption spectrometer used for recording the hyperfine structure of the rovibrational line and will be described in section 3. Figure 1.a represents the linear absorption spectrometer. A Fabry-Perot cavity (FPC) with a 1 GHz free spectral range and a finesse of 150 is then used to drastically filter out the residual carrier and the unwanted *SB+* sideband and to stabilize the intensity of the transmitted sideband *SB-*. In order to keep the laser intensity constant at the entrance of the cell during the whole experiment, the transmitted beam is split in two parts with a 50/50 beamsplitter: one part feeds a 37-cm long ammonia absorption cell for spectroscopy (probe beam *B*) while the other is used as a reference beam (reference beam *A*). The reference signal *A* intensity which gives the intensity of the sideband *SB-* is compared and locked to a very stable voltage reference (stability better than 10 ppm) by acting on the length of the FPC. A suitable intensity discriminator is obtained when the FPC is tuned so that the sideband frequency lies on the slope of the resonance. The absorption length of the cell can be adjusted from 37 cm (in a single pass configuration) to 3.5 m (in a multi-pass configuration). Both the reference beam (*A*) and the probe beam (*B*) which cross the absorption cell are amplitude-modulated at $f_1 = 40$ kHz via the 8-18 GHz EOM for noise filtering and signals are obtained after demodulation at $f_1$. The probe beam (*B*) signal then gives the absorption signal of the molecular gas recorded with a constant incident laser power governed by the stabilization of signal *A*. The sideband is tuned close to the desired molecular resonance and scanned over 250 MHz to record the Doppler profile.

*2.2. The thermostat*
This experiment requires a thermostat to maintain the spectroscopic cell at a homogenous temperature [42]. The absorption cell sits in a large thermostat filled with an ice-water mixture, in order to set its temperature close to 273.15 K. The thermostat is a large stainless steel box $(1.2\times0.8\times0.8\ m^3)$ thermally isolated by a 10-cm thick insulating wall (see Figure 2). The absorption cell $(33\times18\times9\ cm^3)$, placed at the centre of the thermostat, is a stainless steel vacuum chamber ended with two anti-reflective coated ZnSe windows. From these windows, pumped buffer pipes extend out of the thermostat walls. They are closed on the external other side with room temperature ZnSe windows. Vacuum prevents heat conduction and water condensation on windows.

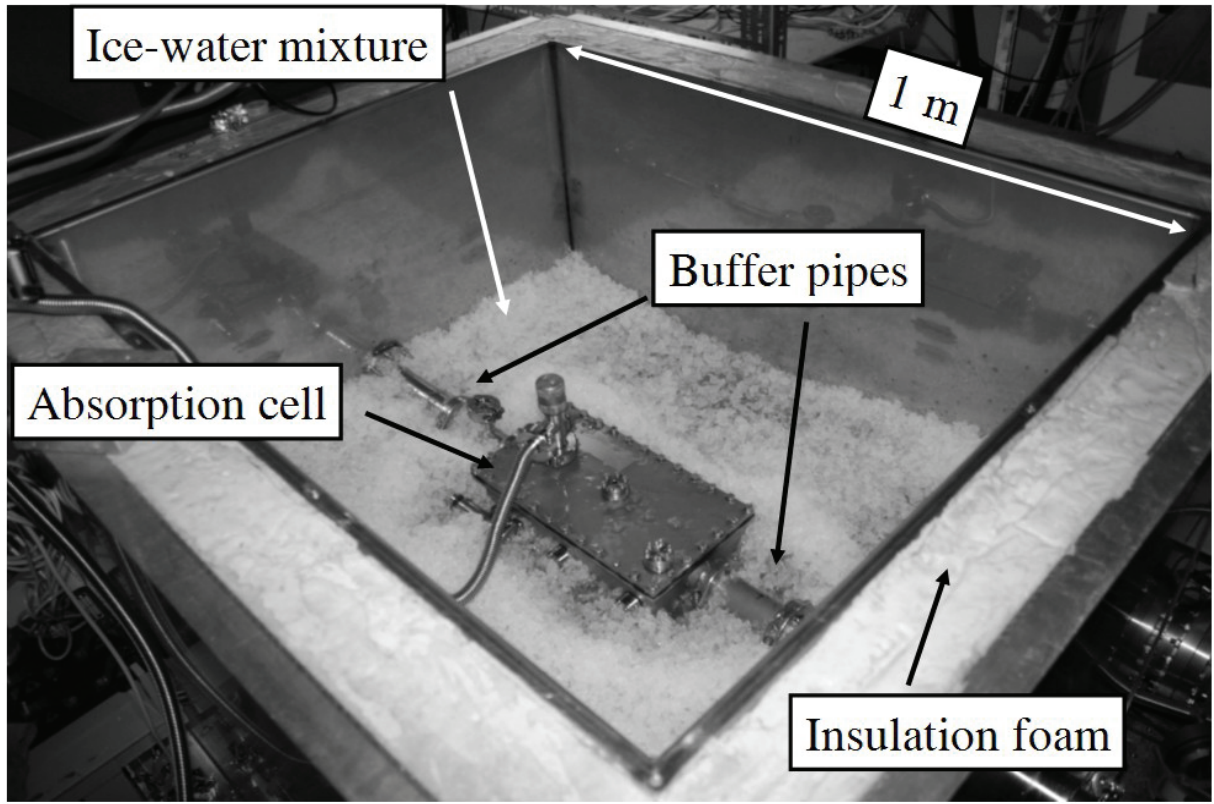

**Figure 2. Absorption cell inside the ice-water thermostat.**

The cell temperature and thermal gradients are measured with long stems 25 Ω Standard Platinum Resistance Thermometers (SPRTs) calibrated at the triple point of water and at the gallium melting point. Those SPRTs are compared to a low temperature dependence resistance standard in an accurate resistance measuring bridge. The resulting temperature accuracy measured close to the cell is 1 ppm with a noise of 0.2 ppm after 40 s of integration. For longer integration times, temperature drifts of the cell remain below 0.2 ppm/h. The melting ice temperature homogeneity close to the cell has been investigated. Reproducible residual gradients parallel to the cell walls have been measured: the vertical – resp. horizontal (both directions) – gradient is equal to 0.05 mK/cm ie 0.17 ppm/cm– resp. 0.03 mK/cm ie 0.1 ppm/cm – leading to an overall temperature inhomogeneity along the cell below 5 ppm. These residual temperature gradients probably come from the difficulty to keep a homogeneous mixture surrounding the cell. Finally, we conclude that the temperature in the experiment is $T$=273.1500 (7) K.

**3. Hyperfine structure of the ammonia line**
The saQ(6,3) line ($J$ = 6 and $K$ = 3 are respectively the quantum numbers associated with the total orbital angular momentum and its projection on the molecular symmetry axis) has been chosen because it is a well-isolated rovibrational line with long-lived levels (natural width of the order of a few Hz). However, owing to the non-zero spin values of the N and H nuclei, an unresolved hyperfine structure is present in the Doppler profile of the rovibrational line and is responsible for a broadening of the line which is related to the relative position and strength of the individual hyperfine components. The relative increase of the linewidth due to this hyperfine structure scales as the square of the ratio $\Delta_{hyp}/\Delta\nu_{Dopp}$ (where $\Delta_{hyp}$ is the global spread of the overall hyperfine structure and $\Delta\nu_D = \Delta\omega_D/2\pi$, the Doppler width) which results in a relatively small influence. In the case of the probed ammonia line, we will see that the hyperfine structure extension of the stronger components is of the order of 50 kHz. However, weaker lines around ±600 *kHz* away from the main structure must be considered as they actually give the largest contribution. For a Doppler width of about 50 MHz, the impact may be a few ppm. For this reason, it is necessary to have a good knowledge of that structure in order to take it into account in the line shape analysis.

*3.1. Description of the hyperfine interactions*
The hyperfine Hamiltonian of ammonia is very well-known [43-45]. The hyperfine structure of the saQ(6,3) line is in part due to the interaction between the nitrogen nuclear quadrupole moment and the gradient of the electric field at the nucleus. Spin-rotation terms come from the interaction between the magnetic field induced by the molecular rotation and the magnetic moment of the nitrogen nucleus and the hydrogen nuclei. The other magnetic hyperfine terms are the spin-spin interactions between N and H atoms or between H atoms themselves. The strength of these interactions is characterized by coupling constants usually noted *eQq* (N quadrupole constant), *R* (N spin-rotation constant), *S* (H spin-rotation constant), *T* (N-H spin-spin constant) and *U* (H-H spin-spin constant) according to notations first introduced by Kukolich [46]. Those constants are experimentally accessible. There are two sets of such constants for the fundamental and the upper rovibrational levels. Since the nitrogen nuclear spin is $I_N$=1, each rovibrational level is split in 3 sub-levels $F_1$ = (7, 6, 5) where $F_1$ is the modulus of the sum of the orbital angular momentum and the spin of the nitrogen nucleus, $\vec{F_1} = \vec{J} + \vec{I_N}$. Then, each of these sub-levels is again split in 4 sub-levels characterized by $(F_1, F)$ where $F = (F_1 \pm 1/2, F_1 \pm 3/2)$ is the modulus of the total angular momentum of the molecule, $\vec{F} = \vec{F_1} + \vec{I}$. $\vec{I}$ is the total spin of the hydrogen nuclei. Its modulus is equal to 3/2 when *K* is a multiple of 3 [43].

*3.2. Saturation spectroscopy*
The first hyperfine structures of ammonia were recorded on a molecular beam in the microwave region [45-47] and led to a very good knowledge of the hyperfine constants in the ground vibrational level. Saturation spectroscopy of ammonia in the infrared leads to extra information on the upper vibrational level and was first performed in our group exhibiting both the electric and magnetic hyperfine structure of ammonia [48, 49], especially the six components of the asQ(8,7) line, fully resolved in a large 18-m long absorption cell [50]. From these measurements the variation of the quadrupole constant and spin-rotation constants with vibration could be obtained. For the present study a new experimental setup has been developed to record the hyperfine structure of the saQ(6,3) rovibrational line by saturated absorption spectroscopy (see figure 1.b). A 3-m long Fabry Perot cavity in a plano-convex configuration with a convex mirror radius of 100 m and a finesse of about 140 is filled with ammonia. The red detuned *SB-* sideband generated by the 8-18 GHz EOM feeds this Fabry-Perot cavity. Two frequency modulations, $f_2$ and $f_3$ are required for this experiment. The modulation $f_2$ is used to stabilize the resonator frequency and can be applied either on one mirror mounted on a piezoelectric transducer or directly on the sideband frequency via the synthesizer which drives the 8-18 GHz EOM. The hyperfine components of the molecular line are detected in transmission of the cavity after demodulation at $f_3$, a modulation applied on the sideband frequency via the EOM. Experimental parameters were first optimized to reduce as much as possible the linewidth in order to clearly observe the three main $\Delta F_1$=0 lines. On the spectrum displayed on Figure 3, those main components are well fitted by first derivatives of Lorentzians. Each line presents an unresolved structure of 4 $\Delta F$=0 components. The modulation applied on the FPC for its frequency stabilisation was at 11 kHz and the sideband modulation frequency for the molecular lines detection was equal to 1 kHz with a depth of 2 kHz. The resolution was limited by laser intensity (around 1 mW inside the Fabry-Perot cavity), gas pressure ($10^{-5}$ mbar), modulation settings and transit time broadening. The absorption signal was recorded over 200 kHz with 500 points and 30 ms integration time per point.

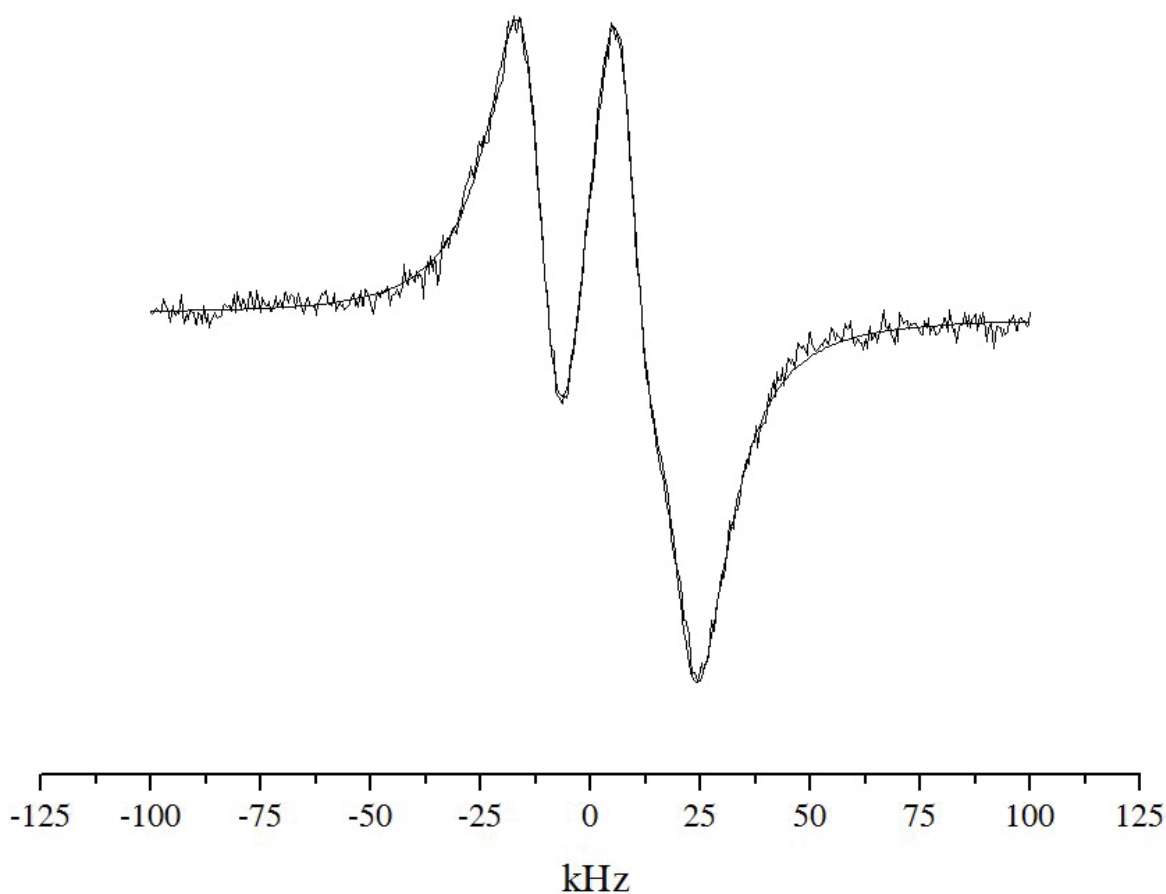

**Figure 3. Main $\Delta F_1 = 0$ components of the $\nu_2$ saQ(6,3) experimental spectrum (first harmonic detection) of $^{14}NH_3$ fitted by three derivatives of a Lorentzian.**

The experimental hyperfine spectrum of Figure 3 has been fitted with three derivatives of a Lorentzian lineshape. The adjustable parameters were the baseline offset and slope, the line central frequency, the intensity scale, the full width at half maximum (FWHM) of the Lorentzian (identical for the three components), $\Delta eQq$ and $\Delta R$, respectively the change in the quadrupole coupling constant and in the N spin-rotation constant between the upper and lower levels. Figure 3 illustrates the excellent agreement between experimental data and the numerical adjustment.

These 12 partially resolved lines are the strongest lines corresponding to an approximate selection rule $\Delta J = \Delta F_1 = \Delta F$. In fact, 66 weaker transitions are also allowed and will contribute to the Doppler signal and broaden it. Doppler-generated level crossing resonances can also be observed in saturated absorption (but are not present in linear absorption spectroscopy) and give signal at the mean frequency between the two involved transitions. Figure 4 compares (a) the $\nu_2$ saQ(6,3) linear absorption signal (recorded over 1 GHz using sideband amplitude modulation at $f_1$=40 kHz and first harmonic detection, see section 2.1) and (b) the saturated absorption signal recorded over 1.4 MHz in transmission of the 3-m long Fabry-Perot cavity. In the latter case, experimental parameters have been adjusted to optimize the signal-to-noise ratio in order to be able to observe the expected weak satellite transitions. The cost to be paid is a degradation of the resolution and a slight distortion of the line shape. All frequency modulations were directly applied on the sideband. A 90 kHz frequency modulation (60 kHz depth) was used for the resonator frequency stabilisation. For molecular line detection a 10 kHz frequency modulation (30 kHz depth) was applied and first harmonic detection was used (with 30 ms integration time per point).

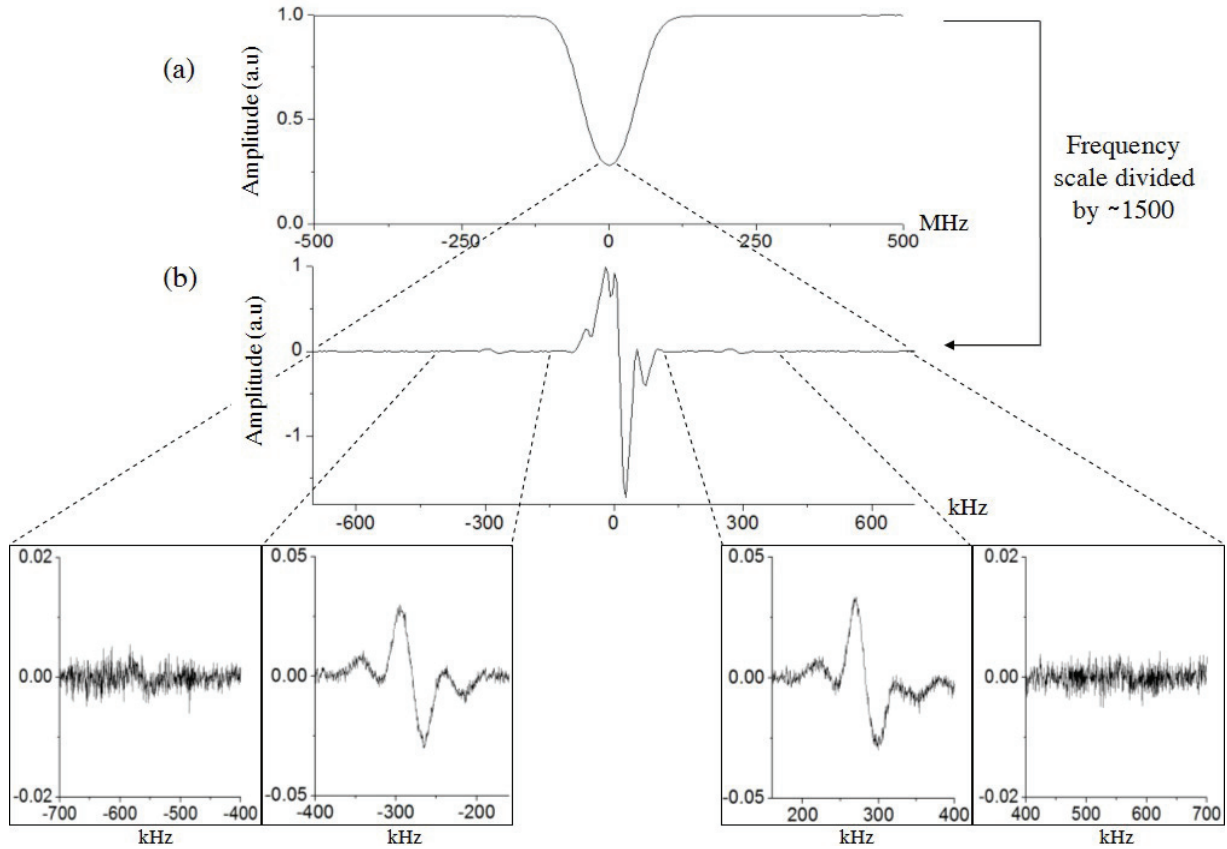

**Figure 4.** $^{14}NH_3$ saQ(6,3) absorption line recorded by linear absorption (a) and at higher resolution by saturated absorption spectroscopy (b). At about 300 kHz on both sides of the central components Doppler-generated level crossings (between $\Delta F=0$ and $\Delta F=\pm 1$) are observed. At about 600 kHz from the central resonances satellite weaker components are expected.

Under these experimental conditions, the three intense $\Delta F_I=0$ multiplets are strongly broadened by the frequency modulation. Signal of Doppler-generated level crossings (between $\Delta F=0$ and $\Delta F=\pm 1$) is clearly observed around $\pm 300\ kHz$ from the central components. For a frequency detuning of about $\pm 600\ kHz$ from the central components, signals coming from very weak $\Delta F=\pm 1$ satellite components and crossovers between $\Delta F=0$, $\Delta F=+1$ and $\Delta F=-1$ are hardly distinguishable.

*3.3. Analysis of the hyperfine structure*
Clearly, the recorded spectra do not allow a determination of the whole set of hyperfine constants. In particular, we can only measure the position of the center of gravity of each series of crossover resonances. However, the numerous studies of hyperfine structures in the ground vibrational level [45-47], allows us to accurately fix the value of the hyperfine constants in the $v=0$ level:
$eQq_0$= -4010 (1) kHz; $R_0$= 6.75 (1) kHz; $S_0$= -18.00 (1) kHz; $T_0$= -0.85 (1) kHz and $U_0$= -2.5 (3) Hz
Only rovibrational saturation spectroscopy provides information on the hyperfine constants in the v=1 level. Our group has recorded in the past the asR(5,0) and asQ(8,7) lines of $^{14}NH_3$ [48, 50] and also the asR(2,0) line of $^{15}NH_3$ [49, 51]. These studies give the right order of magnitude of the hyperfine constants in the upper level of the saQ(6,3) transition. The fit of the three main multiplets (Figure 3) revealed that the uncertainty on their relative positions was 40 Hz and that this structure was only sensitive to the change of $eQq$ and $R$ between the lower and upper levels, leading to:
$\Delta eQq = eQq_1 - eQq_0 = -196.8(6)\ kHz$ and $\Delta R = R_1 - R_0 = -535(6)\ Hz$

The other upper state constants were fixed with a conservative uncertainty of 10% to values estimated from our previous studies on the asR(5,0), asQ(8,7) and asR(2,0) lines:
$S_1 = -17.5(18)\ kHz; T_1 = -0.9(1)\ kHz\ and\ U_1 = -2.5(3)\ Hz$

In principle, the position of the center of gravity of the crossover resonances (with respect to that of the main lines) could give information on the hyperfine structure both in the lower and upper vibrational level. However, our experimental results, with an accuracy of 400 Hz on that position, although in good agreement with the ground vibrational level hyperfine constants, do not bring enough new information.

Figure 5 shows the hyperfine lines as sticks with relative intensities corresponding to the weak field regime. Apart from the strong main lines, the structure reveals manifolds around $\pm 600\ kHz$ (see Figure 4) and $\pm 150\ kHz$ (not investigated by saturated absorption spectroscopy).

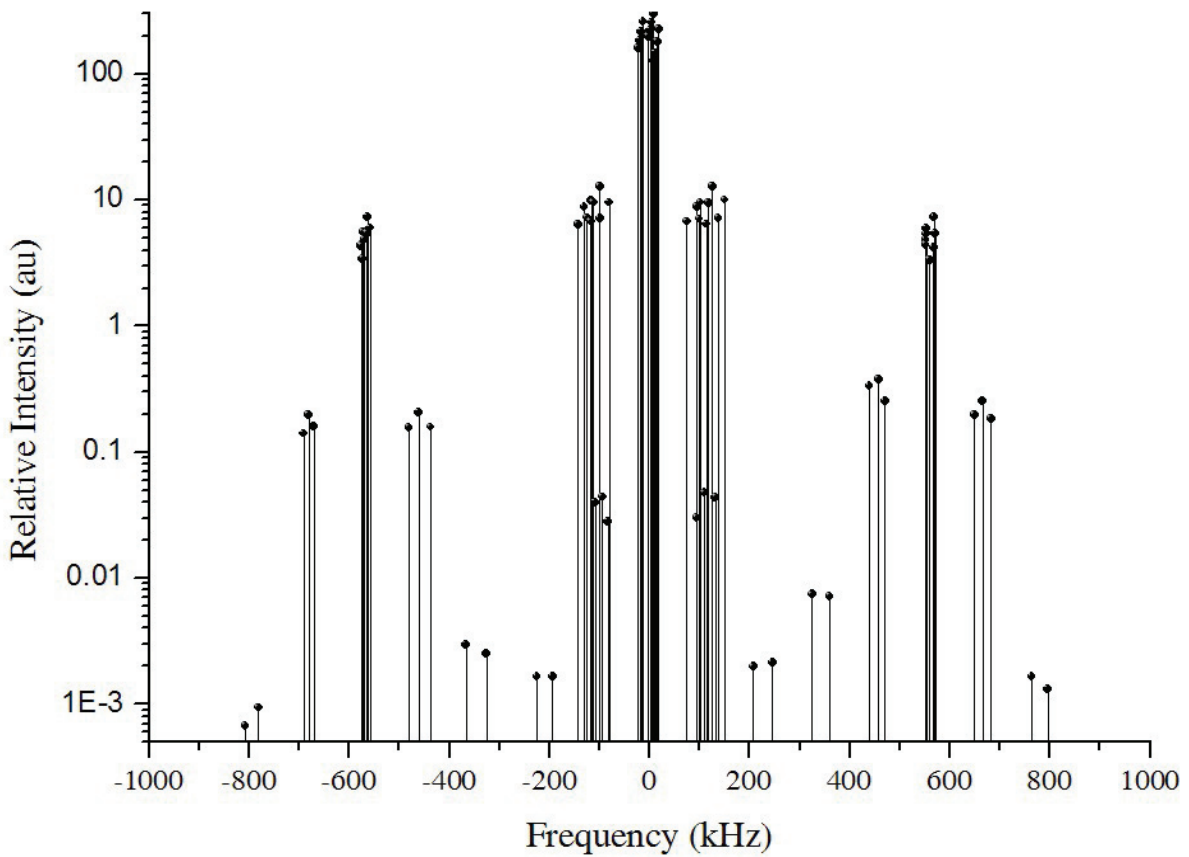

**Figure 5. Stick spectrum of the 78 hyperfine components present in the Doppler profile. The height of each stick reflects the intensity (in logarithm scale) of the corresponding hyperfine transition in linear absorption.**

Using the SPCAT program (developed by H. Pickett, Jet Propulsion Laboratory [52]), as well as a homemade saturation spectroscopy simulation program, we checked very carefully how the positions of the lines and their intensities in linear absorption are affected by a change of the hyperfine constants. This showed that the present knowledge of the hyperfine constants gives a very strong constraint on the hyperfine structure in the Doppler profile of the saQ(6,3), both frequency- and intensity-wise.

As a result of the uncertainty on the values of the hyperfine constants, an uncertainty of 150 Hz on the center of gravity of the crossovers situated around $\pm 300\ kHz$ is deduced. The corresponding uncertainty on the intensities in linear absorption stays below 0.15%. These two effects will fix the uncertainty on the correction due to the hyperfine structure to be applied for the determination of the Doppler width and thus $k_B$.

## 4. The Boltzmann constant measurement

*4.1. Doppler broadening measurement*

*4.1.1. Absorption line shape*

We consider the case of an optically thick medium under low saturation for which the absorption coefficient for the laser power given by $\kappa(\omega) = -\frac{1}{P(z,\omega)} \frac{dP(z,\omega)}{dz}$ (1) leads to the Lambert-Beer law $P(L,\omega) = P(0) e^{-\kappa(\omega)L}$ (for a total absorption length $L$) in the linear regime. At low pressure, the absorption coefficient $\kappa(\omega)$ can be described by a Voigt profile which is the convolution of a Gaussian shape related to the inhomogeneous Doppler broadening and of a Lorentzian shape whose half-width, $\gamma_{ab}$, is the sum of all homogeneous broadening contributions. Since the natural width is negligible for rovibrational levels, this homogeneous width is dominated by molecular collisions and thus, is proportional to pressure. In linear absorption spectroscopy and for an isotropic distribution of molecular velocities it has been recently demonstrated that all transit effects are already included in the inhomogeneous Doppler broadening and do not depend on the laser beam profile, a result which is not intuitive [53]. At high pressure, the Lamb-Dicke-Mössbauer (LDM) effect which results in a reduction of the Doppler width with pressure must be taken into account[54-58]. The absorption coefficient is the Fourier transform of the correlation function of the optical dipole, denoted as $\phi(\tau)$. For the dimensionless absorbance $A(\omega - \omega_{ab}) = \kappa(\omega - \omega_{ab})L$ one finds [53]:

$$A(\omega - \omega_{ab}) = \frac{4\pi \alpha N d_{ab}^2 \omega L e^{\left(-E_a/k_B T\right)}}{Z_{int}} \operatorname{Re} \int_0^{+\infty} \exp(-i\omega\tau) \phi(\tau) d\tau \quad (2)$$

where $\omega$ is the laser angular frequency, $\omega_{ab} = \frac{E_b - E_a}{\hbar}$ the angular frequency of the molecular line ($E_a$ and $E_b$ are the energies of the lower and upper rovibrational levels a and b in interaction with the laser electromagnetic field, $E_a < E_b$), $\alpha = e^2/(4\pi\varepsilon_0 \hbar c)$ the fine structure constant (e is the electron charge), $N$ the density of molecules, $d_{ab} = \mu_{ab}/e$ ($\mu_{ab}$ is the transition moment), and $Z_{int}$ the internal partition function.

Various theoretical models are available in the literature to describe the correlation function of the optical dipole, depending on the assumption made for the type of collisions between molecules [53, 59]. Among them the Galatry profile [55] makes the assumption of so-called "soft" collisions between molecules with the introduction of the diffusion coefficient $D$. The Galatry optical dipole correlation function is then:

$$\phi_G(\tau) = \exp\left[i\omega_{ab}\tau - \gamma_{ab}\tau + \frac{1}{2}\left(\frac{\Delta\omega_D}{\beta_d}\right)^2 \{1 - \beta_d\tau - \exp(-\beta_d\tau)\}\right] \quad (3)$$

where $\Delta\omega_D$ is the half-width of the Doppler profile and $\beta_d = \frac{k_B T}{mD}$ the coefficient of dynamical friction ($m$ is the molecular mass). The Galatry absorbance can then be written using the $_1F_1$ Kummer confluent hypergeometric function:

$$A_G(\omega - \omega_{ab}) = \frac{4\pi\alpha N d_{ab}^2 \omega L e^{(-E_a/k_B T)}}{\Delta\omega_D Z_{int}} \operatorname{Re} \frac{1}{y(\xi)} {}_1F_1\left[1, 1 + \frac{y(\xi)}{\theta}; \frac{1}{2\theta^2}\right] \quad (4)$$

where $\theta = \frac{\beta_d}{\Delta\omega_D}$, $y(\xi) = \frac{1}{2\theta} + \eta - i\xi$ and $\zeta = \xi + i\eta = \frac{(\omega - \omega_{ab}) + i\gamma_{ab}}{\Delta\omega_D}$. The Galatry profile evolves from a Lorentzian shape in the high pressure limit to a Voigt profile at low pressure.

At low pressures (small $\beta_d$), we can use for the absorbance the first-order expansion in $\theta$:

$$A(\omega - \omega_{ab}) = \frac{4\alpha N d_{ab}^2 \omega L e^{\left(-\frac{E_a}{k_B T}\right)}}{\sqrt{\pi}\Delta\omega_D Z_{int}} \left(\operatorname{Re} w(\zeta) + \frac{\theta}{12} \operatorname{Re} w_1(\zeta)\right) \quad (5)$$

where $w(\zeta)$ and $w_1(\zeta)$ can be expressed in terms of the error function $w(\zeta) = e^{-\zeta^2} \operatorname{erfc}(-i\zeta)$ and $w_1(\zeta) = \frac{8}{\sqrt{\pi}}(1 - \zeta^2) + 4i\zeta(3 - \zeta^2)\exp(-\zeta^2)\operatorname{erfc}(-i\zeta)$.

The absorbance presents two terms the first one with $w(\zeta)$ corresponds to the Voigt profile when $\theta$ ie $\beta_d$, tends to zero and the second one with $w_1(\zeta)$ is the LMD correction at first order. The expression (5) turns out to be a very good approximation of the true Galatry profile under our conditions (see below) and has been chosen in the fitting procedure as the reference line shape with the advantage of a much faster computing time.

*4.1.2. Measurement and data processing*

The absorption profile, whose Doppler width $\Delta\nu_D = \frac{\Delta\omega_D}{2\pi}$ (the main contribution to the width in our experimental conditions) is of the order 50 MHz, has been recorded over 250 MHz by steps of 500 kHz with a 30 ms time constant. The time needed to record a single spectrum is about 42 s. For 100% absorption, the signal-to-noise ratio is typically $10^3$. Since the signal-to-noise ratio was not high enough to leave the parameter $\beta_d$ as an adjustable parameter, it was kept proportional to the pressure during the numerical adjustment procedure with a proportionality factor deduced from literature. Following the original Galatry theory (based on S. Chandrasekhar's Brownian motion theory) we used the standard diffusion coefficient, found to be equal to $D_{NH_3}^0 = 0.15 \text{ cm}^2\text{s}^{-1}$ at $P_0 = 1 \text{ atm}$, as measured in ref [60] in a classical transport study. Spectroscopic measurements of this coefficient have been performed for other lines of ammonia by A.S. Pine and co workers [61], leading to an effective value 20% smaller than a direct measurement by diffusion in the case of the $\nu_1$ band of $NH_3$. We actually checked that the results of the fits did not change significantly with such a 20% variation of $\beta_d$. Note that the LDM effect scales as the ratio of wavelength to mean free path. The mean free path between collisions, inversely proportional to the pressure, is related to the diffusion coefficient: $l_m = \sqrt{3m/k_B T} \times D_{NH_3}^0 \times (P_0/P)$. It is then easy to find that $\theta = \sqrt{\frac{3}{8\pi^2} \frac{\lambda}{l_m}}$ where $\frac{\theta}{12}$ appears as a scaling factor of the LDM term in expression (5). In our pressure conditions (from 2.5 down to 0.1 Pa) this scaling factor varies from $6 \times 10^{-5}$ to $2 \times 10^{-6}$.

Even with $\beta_d$ kept constant in the fitting procedure, the signal-to-noise ratio of individual spectra was not high enough to disentangle easily the homogeneous and the Doppler width when using usual fitting algorithms. If we rewrite $\gamma_{ab}$ as $gP$ where $P$ is the pressure, proportional to the amplitude of absorption, $g$ is a collisional parameter, a parameter shared by all spectra whatever the pressure is. Thus, to make the fitting algorithm converge, we decided to adjust $g$ in such a way that it is constrained to a constant value for all the measured spectra. We first guess an initial realistic value. We fit all the experimental spectra with a Galatry profile, constraining $g$ to its guessed value, leaving only $\Delta \nu_D = \frac{\Delta \omega_D}{2\pi}$, $P$ (both in the amplitude and $\gamma_{ab}$), $\nu_{ab}$, and the baseline as adjustable parameters. We expect $\Delta \nu_D$ to remain constant when the pressure varies, if $g$ is chosen equal to the correct value. We then plot $\Delta \nu_D$ as a function of $P$ and record the slope $s$ given by a linear regression of this data. We repeat this procedure for different values of $g$ leading both to negative and positive slopes and compute its estimated final value for which $\Delta \nu_D$ remains constant (within the noise) when the pressure varies. The experimental data are finally fitted again constraining $g$ to this final value. A weighted average of all the $\Delta \nu_D$ gives the best estimate of the Doppler width from which we deduce the Boltzmann constant (see ref [62] for more details on the fitting procedure).

*4.2. Statistical uncertainty analysis*

*4.2.1 First series of experiments*
After 16 hours of accumulation, 1420 spectra recorded at various pressures (from 0.1 Pa to 1.3 Pa) yielded a statistical uncertainty on $k_B$ of 37 ppm, limited by noise detection. The statistical uncertainty was calculated as the weighted standard deviation deduced from the dispersion of the 1420 Doppler linewidth measurements [42, 62]. Weights were obtained as the inverse of the square of error bars deduced from the adjustment of each spectrum. Note that those error bars are about 5 times smaller than the standard deviation estimated from the total dispersion of the 1420 measurements. We also estimated the error bar on the Doppler width of each spectrum from a computer based Bootstrap method [63]. The error bar obtained by this method is compatible within ±5% with the error bar obtained from the fitting procedure. The discrepancy observed between the Doppler width standard deviation estimated from the dispersion of the 1420 spectra and the error bar of single Doppler width measurements confirmed by these methods has been attributed to slow drifts of the optical alignement of the laser beam in the absorption cell. Indeed the $CO_2$ laser based spectrometer and the thermostated cell were located on two separate breadboards. Long term drifts of the optical alignement induce slow variations of the amplitude of residual interference fringes on the optical path which are the main source of the baseline instability. This low frequency effect is not observable on each individual spectrum but could affect the global dispersion of repeated measurements over a few hours.

*4.2.2. New optical arrangement*
To overcome this long term instability, the thermostat and the spectrometer have been placed on a single optical table. Better optical alignement stability combined with improvement in the optical isolation and spatial filtering of the laser beam led to an efficient reduction and control over several days of the residual interference fringes. To reduce statistical uncertainty we also chose to increase the molecular absorbance $\kappa(\omega - \omega_{ab})L$ by recording spectra at higher pressures. A second series of 7171 spectra has been recorded and fitted for pressures up to 2.5 Pa. A typical absorption line fitted with the exponential of a Galatry profile and normalized residuals are reported in Figure 6.

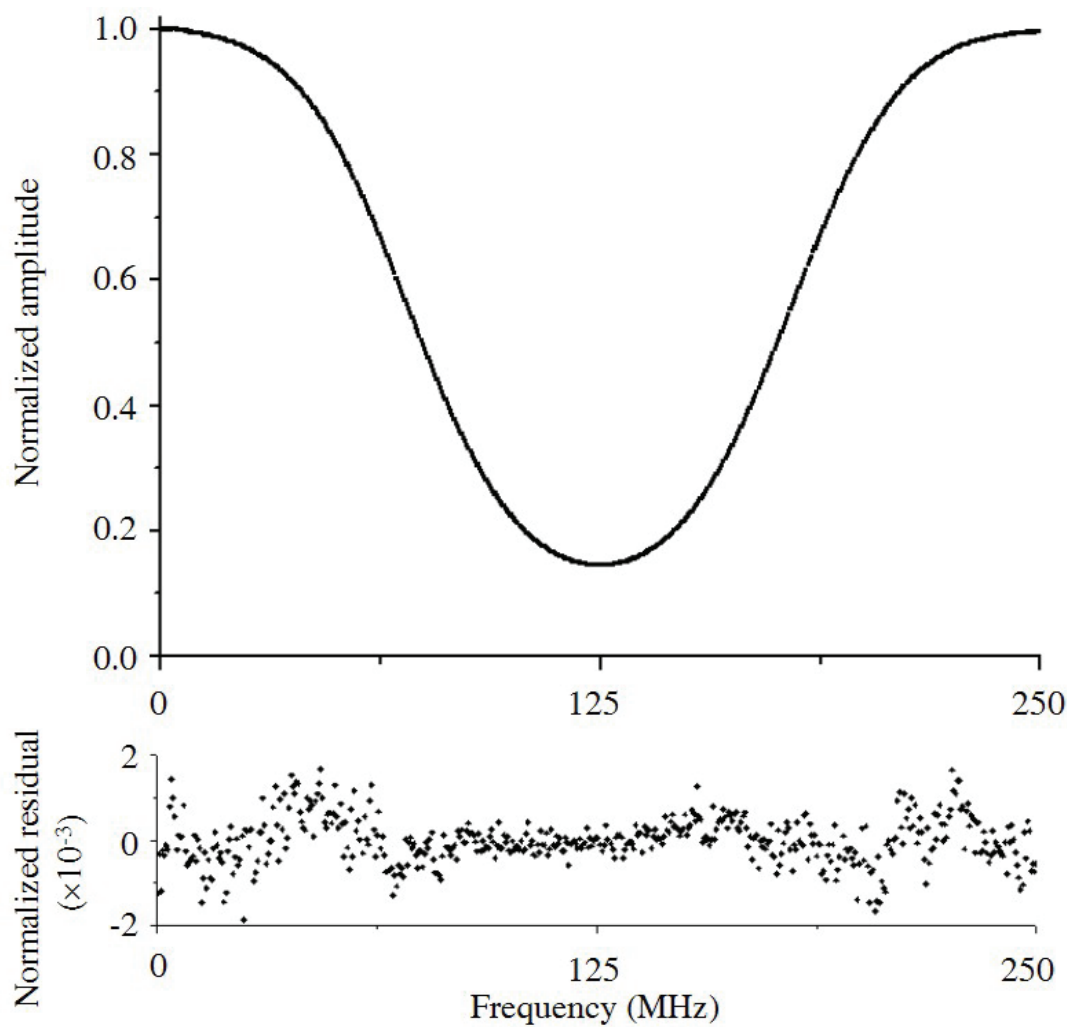

**Figure 6. Absorption spectrum recorded at 1.3 Pa and normalized residuals of a non-linear least-squares fit with the exponential of a Galatry profile Taylor expansion to first order in β$_d$.**

In addition, to take into better account the characteristics of the spectra, a weight of each individual point is attributed by considering the local noise of the spectrum – this is directly related to the intensity noise on the photodetector which decreases strongly when the absorption changes from 0 to about 100%. The values of the Doppler width of the 7171 spectra recorded over 70 hours are displayed on Figure 7 and led to a mean Doppler half-width of: $\Delta \nu_D = 49.88590\,(16)\ MHz\ (3.2\ ppm)$ leading to a statistical uncertainty on the Boltzmann constant determination of 6.4 ppm (Figure 9 and Figure 9).

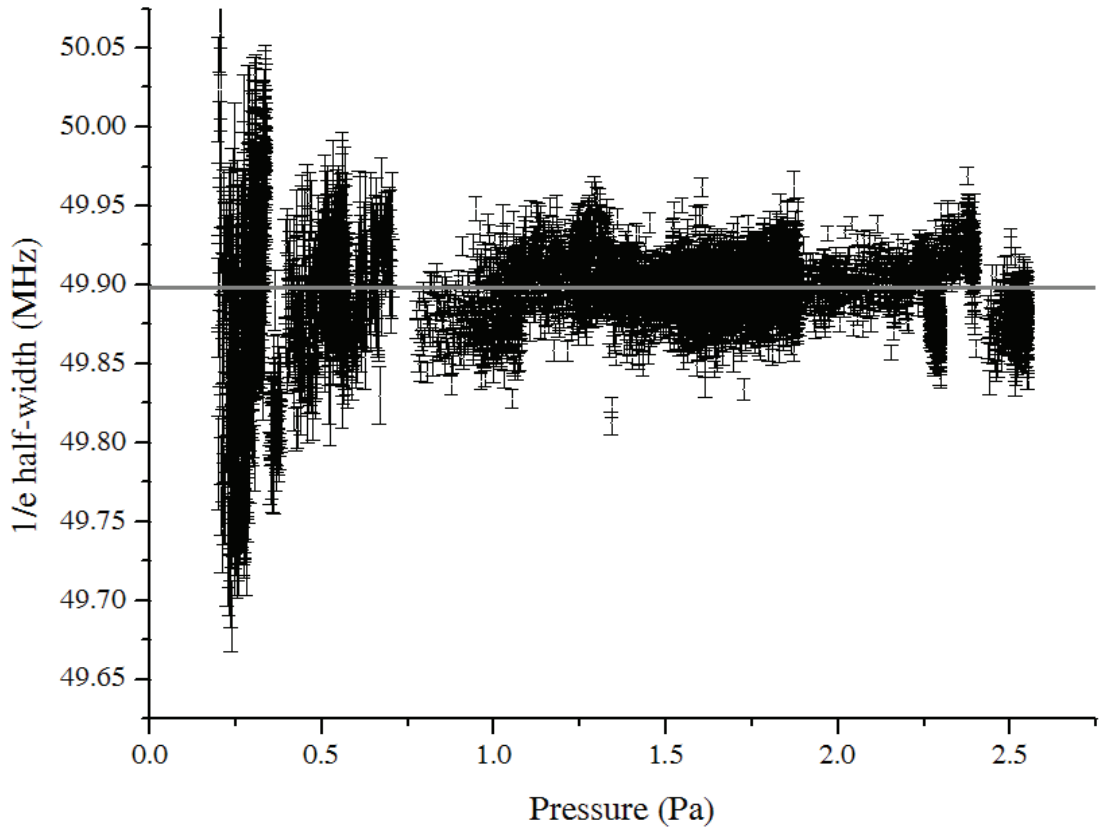

**Figure 7. Doppler e-fold half-width of the saQ(6,3) NH$_3$ absorption line versus pressure, after fitting 7171 spectra with a Galatry profile Taylor expansion to first order in $\beta_d$.**

Note that there is still a discrepancy between the Doppler width uncertainty estimated from measurements dispersion and the error bar on each point estimated either from a non linear regression or the Bootstrap method, but thanks to the improved long term stability of the optical alignement, this discrepancy has been reduced by a factor of 2. The 6.4 ppm error bar reflects dispersion of measurements which includes *de facto* the statistical uncertainty of individual measurements and instabilities of the experiment. Two analyses have been performed to validate this statistical limitation. We randomly divided our data set in four equal subsets of points (each randomly ordered) and analyzed those sets independently to obtain four independant mean values of the Doppler e-fold half-width. The dispersion of these four values reported in Figure 8 is consistent with the statistical uncertainty of each data subset (twice larger than the 3.2 ppm obtained for 7171 spectra).

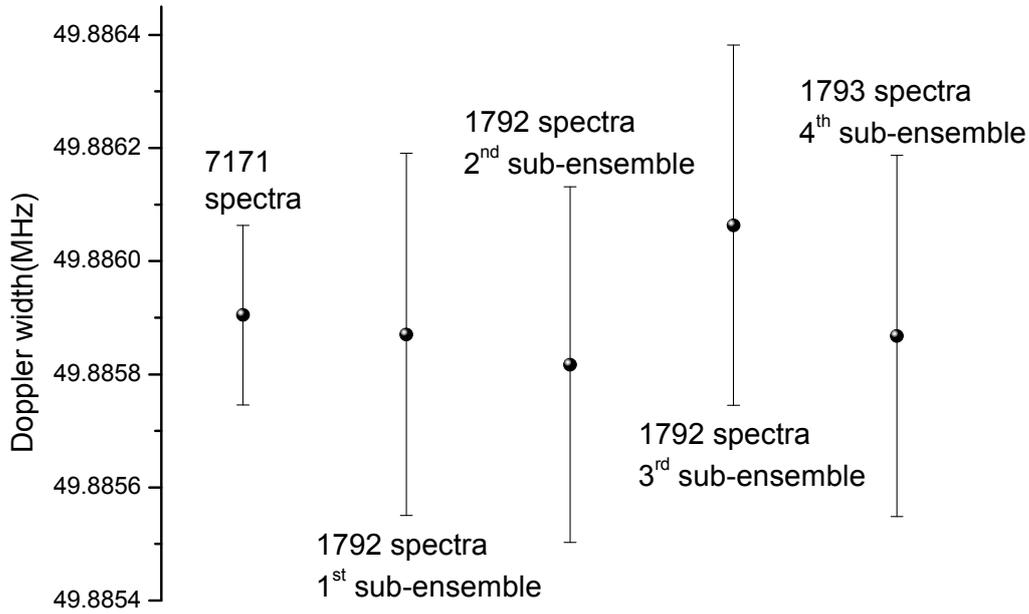

**Figure 8. Mean Doppler e-fold half-width and associated uncertainties for the 7171 spectra and for four subsets of these data (3 sets of 1792 points and 1 set of 1793 points) each randomly ordered.**

In a second analysis, measurements of $k_B$ have been randomly ordered and the relative uncertainty has been calculated as a function of $\tau$, where $\tau$ is the accumulated time of measurement (Figure 9). The slope of both the red and black curve is proportional to one over the square root of $\tau$. This slope is a good indication of a statistical limitation. The significant improvement in the standard deviation at fixed accumulation time is the conjunction of the better optical stability of the setup, the larger pressure range and the new statistical analysis which also clearly stabilizes the fitting procedure and reduces the dispersion of the data.

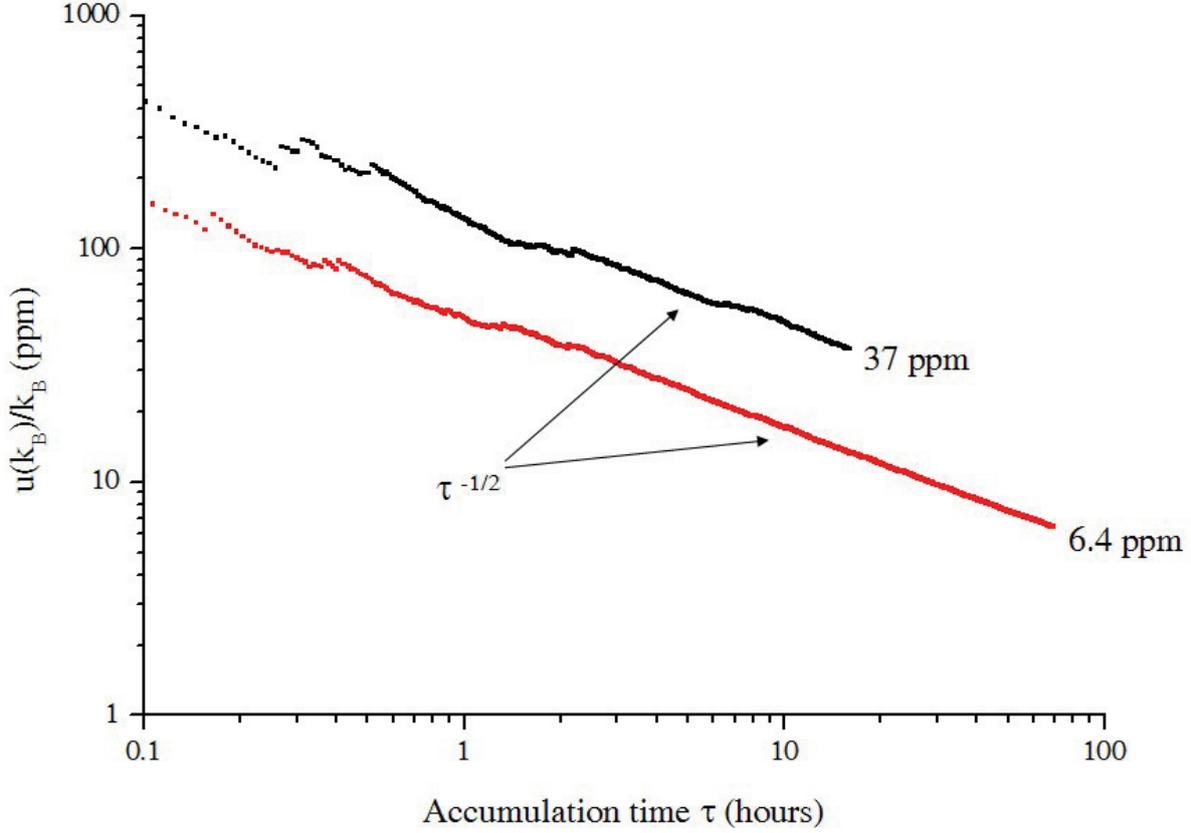

**Figure 9. Relative statistical uncertainty u($k_B$)/$k_B$ of the Boltzmann constant measurement versus time for 1420 spectra recorded over 16 hours and 7171 spectra recorded over 70 hours leading respectively to statistical uncertainties of 37 ppm and 6.4 ppm.**

*4.3. Systematic uncertainty analysis*

The $k_B$ measurement can be affected by several systematic effects. In this section we have listed and investigated some of them: the hyperfine structure of the saQ(6,3) absorption line, the collisional effects, the modulation, the size or the shape of the laser beam, the temperature control of the absorption cell, the non-linearity in the photodetector response and the saturation of the rovibrational transition.

*4.3.1. The hyperfine structure*

Saturated absorption spectroscopy in complement with microwave spectroscopy has provided an accurate determination of spectroscopic parameters of the $\nu_2$ saQ(6,3) line (see section 3). The impact of this hyperfine structure on the Doppler width measurement could finally be estimated. The method is straightforward: we sum Voigt profiles (or Galatry profiles) associated with the 78 hyperfine components of the linear spectrum with positions and intensities precisely determined by the analysis presented in section 3. The resulting lineshape is then fitted by a unique Voigt (or Galatry) profile and the difference with the "true" Doppler width is thus deduced. The 300 Hz uncertainty on the global spread of the overall hyperfine structure (twice that on the crossover positions) and the 0.15% uncertainty on the intensities result in a very precise determination of the correction to be applied on the value of $k_B$: $\left.\frac{\Delta k_B}{k_B}\right|_{hyp.struct.} = -8.71(3)\, ppm$, where 91% of the broadening comes from the weak components around $\pm 600\, kHz$.

This precise evaluation does not take into account any possible differential saturation of the absorption between strong and weak hyperfine transitions. If any, the saturation will be much more important for strong lines which would result in higher relative intensities of the weak lines and thus an additional broadening of the whole Doppler envelope. In order to evaluate this effect, we recorded the Doppler signal and alternatively measured the relative absorption at two different laser powers (0.3 and 0.9 µW) by using optical attenuators either placed just before the photodetector or just before the absorption cell, in order to test saturation effects for a constant detected laser intensity. The absorptions were equal within $5 \times 10^{-4}$ which gives an upper limit for the saturation parameter of $3.6 \times 10^{-3}$ at 1.3 Pa. This very small value is in good agreement with that expected with such laser powers, gas pressure and a typical size of the laser beam of a few mm. In the collisional regime, the saturation parameter scales as the inverse of the square of the pressure. The associated relative broadening stays below 0.3 ppm in the 0.25-2.5 Pa pressure range. Finally it has also been checked that the choice of the individual lineshape (Voigt or Galatry) does not affect the correction on $k_B$.

*4.3.2. The collisional effects*

The 7171 spectra were fitted both with a Galatry (first order Taylor expansion) and a Voigt profile. The relative pressure broadening varies from $6.2 \times 10^{-4}$ to $6.2 \times 10^{-3}$ in the 0.25-2.5 Pa pressure range. A difference of 139 ppm is obtained on $k_B$ when fits of the 7171 spectra with either a Taylor expansion of a Galatry profile or a Voigt profile are compared and this reflects the influence of the LDM effect at high pressure (remember that high pressure data have a stronger weight because of the better signal-to-noise ratio). This illustrates the critical role of the chosen line shape. Each point of Figure 10 shows the fractional difference obtained with the sub-ensemble of these 7171 spectra recorded at pressures below a given $P_{max}$. Figure 10 indicates that the LDM effect is responsible for a narrowing of the Voigt profile leading to differences in the determination of $k_B$ equal or larger than the current statistical uncertainty of 6.4 ppm (see section 4.2.2) for pressure ranges larger than 0.5 Pa. Thus, recording spectra at pressures lower than 0.5 Pa would maintain the systematic error due to the LDM effect below 6.4 ppm, when using a Voigt profile. In addition the quadratic dependence of this difference let us hope to rapidly reduce this effect at the level of 1 ppm.

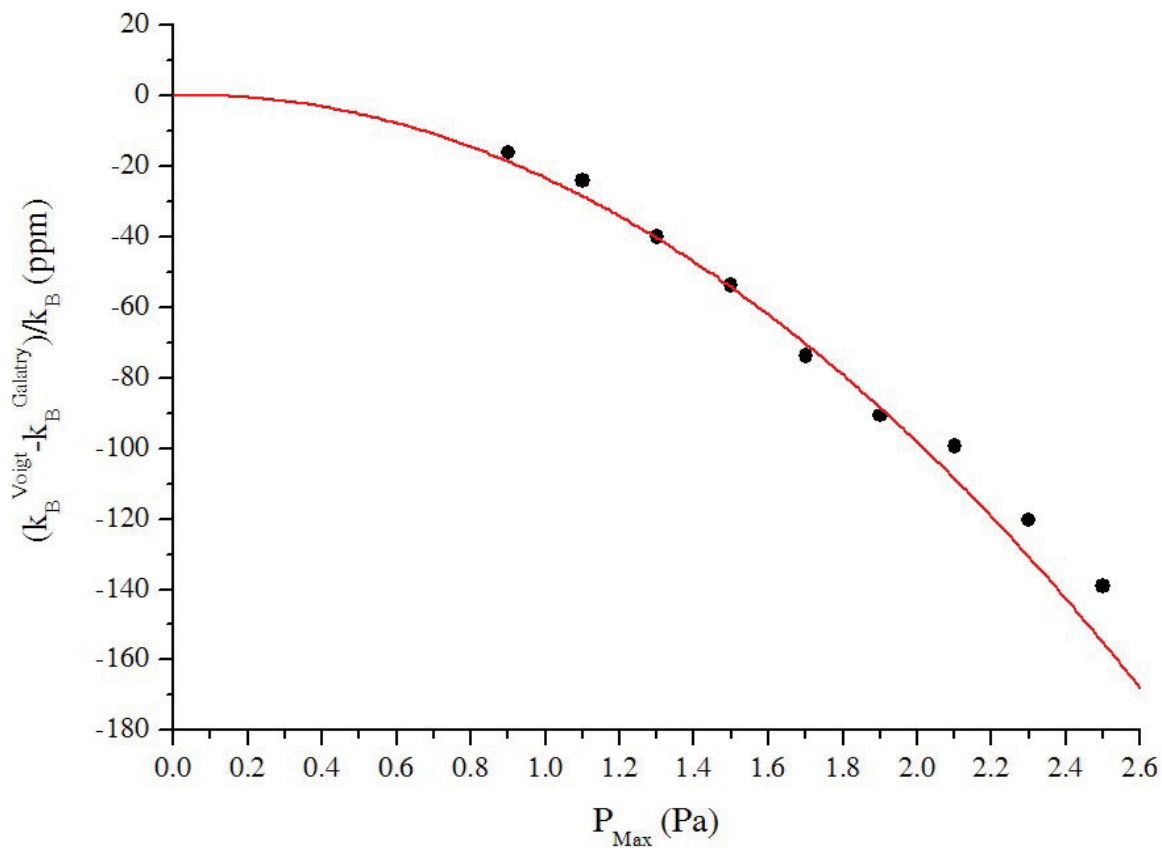

**Figure 10. relative difference on $k_B$ determination (dots) by fitting experimental spectra with a Voigt ($k_B^{\text{Voigt}}$) or a Galatry ($k_B^{\text{Galatry}}$) profile when using the sub-ensemble of the spectra recorded at pressures below $P_{\text{max}}$. The solid line indicates the result obtained when fitting a simulated Galatry profile with a Voigt profile.**

Apart from the Galatry profile, various theoretical models are available in the literature, depending on the assumption made for the type of collisions between molecules [59]. The systematic effect due to the "soft" collisions model chosen here to describe the LDM narrowing would need to be evaluated in our pressure range, by fitting data with other models which would require a precise knowledge of the specific collision kernel. In our experimental conditions (as mentionned in section 4.1.2), the pressure broadening $\gamma_{ab}$ cannot be directly fitted for each individual spectrum but is obtained by adjusting a unique $g$, constrained to a constant value for all the spectra. Thanks to new experimental developments, we are now able to record accurate scans over 500 MHz. This will allow us in the near future to directly and accurately determine this homogeneous broadening for each individual spectrum. In particular, by this way the possible contribution of residual impurities in the absorption cell could be taken into account. This will also permit a more precise study of different line shape models. However we expect the present study to give the right order of magnitude of the LDM narrowing contribution to the determination of $k_B$, whatever the chosen collisional model.

*4.3.3. Other systematic effects*
Attempts to observe other systematic effects due to the modulation and the geometry of the laser beam were unsuccessful at a 10 ppm level. Let us remind that it has been shown theoretically that the line shape does not depend on the laser beam geometry [53].

Taking into account both temperature inhomogeneity and stability (detailed in section 2.2) of the thermostat, no systematic effect due to the temperature control is expected on $k_B$ at a 2.5 ppm level.

Laser power related systematic effects due to both non-linearity in the detection set-up and Saturation broadening of the molecular absorption were investigated too. It is worth reminding that the saturation of the photodetector occurs above 1 mW while the operating conditions are below 1 μW. Boltzmann constant measurements were performed at different saturation parameters (for laser power ranging from 0.5 to 1μW at the entrance of the absorption cell). Non-linearity in the photodetection response was evaluated by recording spectra and determining $k_B$ at different detected powers using attenuators placed straight before the photodetector, in order to work at constant molecular transition saturation. No systematic effects were observed at a 10 ppm level for these two potential causes of systematic effect

In the following table, are summarized the various contributions to the linewidth with their present uncertainty which are systematic effects to be taken into account for a proper evaluation of the Doppler width. In fact, for several non-observable effects only an upper limit estimated from the signal-to-noise ratio can be given. Let us remind that the uncertainties must be doubled when the error budget of $k_B$ is concerned.

**Table 1. Error budget on the determination of the linewidth in parts per million (systematic effects and relative uncertainty).**

| Effect | Relative contribution to the linewidth | Uncertainty |
|---|---|---|
| Doppler width (49.883 MHz) @ 273.15 K | $10^6$ | 3.2 |
| Collisional effects (LMD effect and homogeneous width, for 0.25-2.5 Pa pressure range) | $6.2 \times 10^3$ @ 2.5 Pa | 70 |
| Hyperfine structure of the absorption line | 4.355 | 0.015 |
| Gas purity (partial pressure of impurities from outgasing) | < 10 | < 10 |
| Non-linearity of the photodetector | < 10 | < 10 |
| Saturation broadening of the absorption (for 0.25-2.5 Pa pressure range) | < 10 | < 10 |
| Residual optical offset (from simulations) | < 1 | < 1 |
| Amplitude modulation @ 40 kHz (from simulations) | 0.8 | 0.08 |
| Differential saturation of the hyperfine components (@ 0.9 μW and 0.25-2.5 Pa pressure range) | < 0.3 | < 0.3 |
| Laser linewidth | < 0.2 | < 0.2 |
| Temperature of the gas | 0 | 1.25 |
| Linearity and accuracy of the laser frequency scale | 0 | < 0.01 |
| Transit effect (laser beam geometry) | 0 | 0 |

These figures are usefully compared to the present statistical uncertainty of 3.2 ppm on the Doppler width. Clearly the line shape model is up to now the critical point in this experiment. However, we are confident that the next generation of experiments will lead to a better control of this effect, thanks to an operation within a ten times lower pressure range and to a more accurate fit of individual spectra.

The value of the Boltzmann constant deduced from these 7171 points, corrected for the systematic effect due to the hyperfine structure (see section 4.3.1.), is $k_B = 1.38080(20) \times 10^{-23} \text{J.K}^{-1}$. The combined standard uncertainty is 144 ppm and this value of $k_B$ deviates from that recommended in 2007 by the CODATA by 106 ppm [15].

## 5. Conclusion and perspectives

Recent experimental developments to reduce the statistical uncertainty on the Boltzmann constant determined by linear absorption of ammonia around 10 μm have been reported. New measurements of the Doppler broadening of the $\nu_2$ saQ(6,3) rovibrational line of ammonia close to 273.15 K led to a statistical uncertainty of 6.4 ppm on $k_B$ (after a cumulative measurement time of 70 hours). The improvement on the statistical uncertainty of the Boltzmann constant measurement throughout time at the Laboratoire de Physique des Lasers as well as results from other groups [31,32,35] are illustrated in Figure 11.

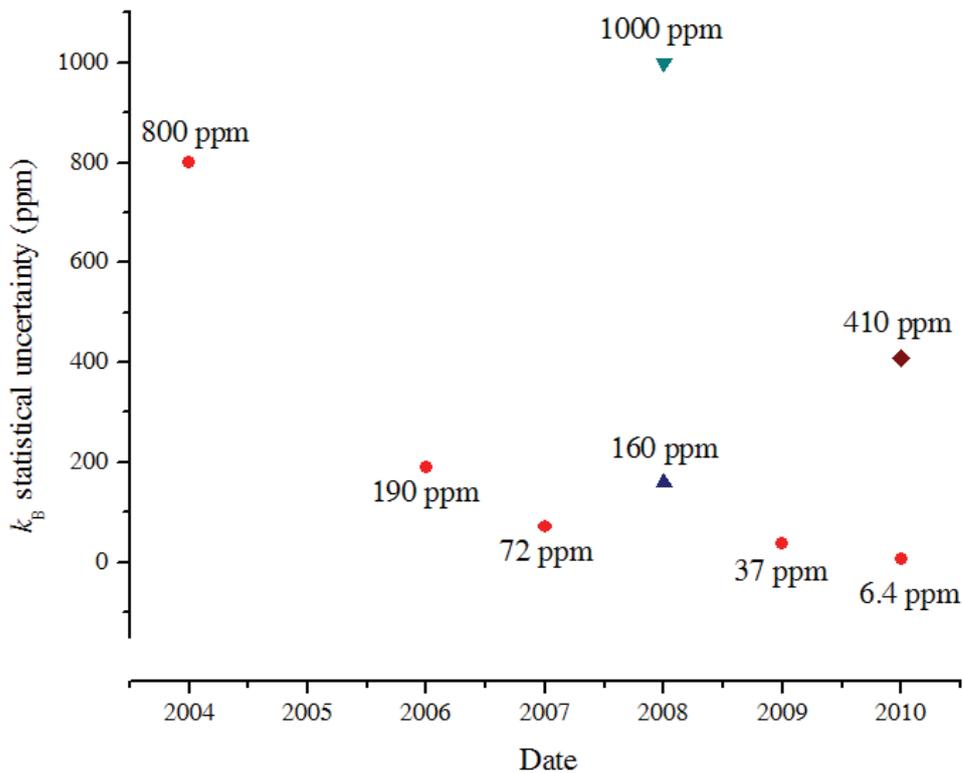

**Figure 11. Statistical uncertainty on the Boltzmann constant measured by laser spectroscopy : (●) at LPL since the first experiments in 2004 [28, 29, 40, 42, 62], (▼) from Yamada *et al.* [32], (▲) from Casa *et al.* [31] and (♦) from Truong *et al.* [35].**

This paper also presents a careful study of the influence of the hyperfine structure. Its role cannot be neglected at a few ppm level but it can now be easily taken into account with a negligible residual uncertainty.

The determination of $k_B$ by laser spectroscopy is affected by several systematic effects. Some of them have been investigated leading to a provisory conclusion that the main source of systematic effects comes from collisional effects. Their contribution to the linewidth is clearly model-dependent. Presently the most accurate value for the Boltzmann constant that can be deduced from Doppler spectroscopy at LPL is the one obtained from measurements performed at very low pressure, below 1.3 Pa (see section 4.2.1.). From the present paper, a correction due to the hyperfine structure of the probed rovibrational line needs to be applied. This correction is of -8.71(3) ppm leading to a refined determination of $k_B = 1.380704(69) \times 10^{-23}$ J.K$^{-1}$. The combined standard uncertainty is dominated by two terms [42]: the statistical uncertainty of 37 ppm and the collisional effects modelisation systematic uncertainty of 34 ppm. This measurement of $k_B$ is in agreement with the value recommended in 2007 by the CODATA, $1.3806504(24) \times 10^{-23}$ J.K$^{-1}$, within 39 ppm [15].

The detailed comparison between the Voigt and the Galatry profile convinced us that the residual error will be reduced at the ppm level if the pressure in the absorption chamber stays below 0.5 Pa. This major conclusion motivates the next evolution of the experimental setup. A new multi-pass Herriott cell (37 m absorption length) will be installed to record spectra at pressures ten times lower than in the experiment presented here while keeping the same signal-to-noise ratio. In this new pressure range (0.025-0.25 Pa) we expect to control the collisional effects at the ppm level. In addition, a new thermostat in which the absorption cell is surrounded by a copper thermal shield located itself in a stainless steel enclosure has been recently developped and installed. A cell temperature inhomogeneity and stability of 1 ppm over 1 day has been already demonstrated [42]. To enable a variable working temperature, useful for a complete analysis of temperature dependent systematic effects, the melting ice will be replaced by a 1 m$^3$ mixture of water and alcohol, maintained at a desired temperature. A cryostat actively coupled to a heat exchanger will permit a regulation of the liquid bath temperature from +10°C down to –10°C. In this temperature range, the relative uncertainty associated with the interpolated temperature measured by the SPRTs used remains within 1 ppm. Finally, we plan to perform a more refined study of the possible influence of the modulation distortion connected to the non-linearity of the photodetector.

These future developments should lead to a new value for the Boltzmann constant with an accuracy of a few ppm which is the main goal of this project. This will be worthily compared to the value obtained by the acoustic method and thus hopefully contribute significantly to the CODATA value. The determination of the Boltzmann constant with different methods at a few ppm is a prerequisite for a new definition of the Kelvin by fixing the value of $k_B$.

**Acknowledgements**

This work is funded by the Laboratoire National de Métrologie et d'Essais and by European Community (EraNet/IMERA). Authors would like to thank A. Amy-Klein for her reading of the manuscript, Y. Hermier, F. Sparasci and L. Pitre from Laboratoire Commun de Métrologie LNE-CNAM for SPRTs calibrations, discussions and advices on the thermometry side of this project, and S. Briaudeau for helping in the design and setting up of the thermostat and temperature control devices.